\documentstyle[aps,prd,twocolumn,tighten]{revtex}
\begin{document}
\newcommand{\Fstar}{\!\stackrel{*}{F}{\!\!}}
\newtheorem{lemma}{Lemma}
\renewcommand{\thelemma}{\Alph{lemma}}
\preprint{ICI-EFEI-002/99}
\draft
\title{Geometrical aspects of light propagation 
in nonlinear electrodynamics}
\author{M. Novello$^1$%
\thanks{Electronic mail: \tt novello@lafex.cbpf.br},  
V. A. De Lorenci$^2$%
\thanks{Electronic mail: \tt lorenci@cpd.efei.br}, 
J. M. Salim$^1$ and R. Klippert$^{1,2}$}
\address{\mbox{}\\$^1$Centro Brasileiro de Pesquisas F\'{\i}sicas \\
Rua Dr.\ Xavier Sigaud 150, Urca 22290-180 Rio de Janeiro, RJ -- Brazil} 
\address{$^2$Instituto de Ci\^encias -- 
Escola Federal de Engenharia de Itajub\'a \\
Av.\ BPS 1303 Pinheirinho, 37500-000 Itajub\'a, MG -- Brazil}
\date{\today}
\renewcommand{\thefootnote}{\fnsymbol{footnote}}
\twocolumn[
\hsize\textwidth\columnwidth\hsize\csname@twocolumnfalse\endcsname 
\maketitle

\begin{abstract}
\hfill{\small\bf Abstract}\hfill\smallskip
\par
We analyze the propagation of light 
in the context of nonlinear electrodynamics, 
as it occurs in modified QED vacua.  
We show that the corresponding characteristic equation 
can be described in terms of a modification of the effective geometry 
of the underlying spacetime structure. 
We present the general form for this effective geometry and 
exhibit some new consequences that result from such approach.
\end{abstract}
\pacs{PACS numbers:  12.20.Ds, 11.10.Wx, 41.20.Jb}
\smallskip\mbox{}]
\footnotetext[1]{Electronic mail: \tt novello@lafex.cbpf.br}
\footnotetext[2]{Electronic mail: \tt lorenci@cpd.efei.br}
\renewcommand{\thefootnote}{\arabic{footnote}}

\section{Introduction}

\subsection{Introductory remarks}
Modifications of light propagation in different vacua states 
has recently been a subject of interest. Such investigation shows 
that, under distinct non trivial vacua (related to several 
circumstances such as temperature effects, 
particular boundary conditions, quantum polarization, etc), 
the motion of light can be viewed as  electromagnetic waves 
propagating through a classical dispersive medium.  
The medium induces modifications on the equations of motion, 
which are described in terms of nonlinearities of the field. 
In order to apply such a {\em medium interpretation} 
we consider modifications of electrodynamics 
due to virtual pair creation.  
In this case the effects can be simulated by an effective Lagrangian 
which depends only on the two gauge invariants $F$ and $G$ 
of the electromagnetic field \cite{Dittrich,Shore}.  

One of the main achievements of such investigation 
is the understanding that, in such nonlinear framework, 
{\it photons} propagate along geodesics that are no more null in 
the actual Minkowski spacetime but in another effective geometry.  
Although the basic understanding of this fact --- 
at least for the specific case of Born-Infeld electrodynamics  --- 
has been known for a long time \cite{Plebansky}, 
it has been scarcely noticed in the literature.  
Moreover, its consequences were not exploited any further.  
In particular, we emphasize the general application 
and the corresponding consequences 
of the method of the effective geometry outlined here.  

The exam of the photon propagation beyond Maxwell electrodynamics 
has a rather diversified history: it has been investigated 
in curved spacetime, as a consequence of non-minimal coupling 
of electrodynamics with gravity \cite{Drumond,Novello} 
and in nontrivial QED vacua, as an effective modification 
induced by quantum fluctuations \cite{Latorre,Dittrich,Shore}. 
As a consequence of this examination some unexpected results appear.  
Just to point one out, we mention the possibility 
of {\em faster-than-light} photons%
\footnote{The meaning of such expression is that 
the wave discontinuities propagate along spacelike 
characteristic surfaces in the Minkowski background.  }. 

The general approach of all these theories is based on a gauge invariant 
effective action, which takes into account modifications 
of Maxwell electrodynamics induced by different sorts of processes.  
Such a procedure is intended to deal with the quantum vacuum 
as if it were a classical medium.  Another important consequence 
of such point of view is the possibility to interpret all such vacua 
modifications --- with respect to the photon propagation --- 
as an effective change of the spacetime metric properties.  
This result allows one to appeal to an analogy with the 
electromagnetic wave propagation in curved spacetime 
due to gravitational phenomena. 

\subsection{Synopsis}

In this paper we deal with light propagation in nonlinear 
electrodynamics. The origin of such nonlinearity 
is not unique. It can be a consequence of modified QED vacua
\cite{adler,scham,barton} or deal to nonlinear response of 
a dielectric medium.  There are different 
ways to evaluate the characteristic surfaces of the wave propagation.  
However, there seems to be  no better and elegant manner 
than the one proposed by Hadamard.  
We briefly review its main lines in Section \ref{geometry}.  
Firstly, we deal only with one-parameter Lagrangians, which 
means that those theories depend only upon the gauge invariant $F$.  
In the subsequent section we generalize for the full dependence 
upon the two algebraic invariants of electrodynamics.  
We show an elegant property of such nonlinear theories: 
the electromagnetic wave propagation can be described 
as if the metric structure of the background were changed 
from its Minkowskian value into another effective metric, 
which depends on the dynamics 
of the background electromagnetic field.  Thus, 
this equivalence property mimics the corresponding properties 
of the photon propagation in gravitational fields.  
Indeed, as we will show in Section \ref{consequences}, photons 
described by nonlinear electrodynamics propagate as null geodesics 
in an effective metric that is distinct from the Minkowskian one.  
We show a remarkable consequence of such interpretation: 
the possibility of generating a compact domain 
in which photons are trapped by the nonlinear electromagnetic field.  
This suggests the possibility that an analogy 
with a gravitational black hole 
-- which we should name electromagnetic black hole -- could exist.  
Besides, we show that one can find a similar phenomenon inside a 
dielectric medium that responds nonlinearly to an external stimulus.  
In Section \ref{vacua} we analyze the light velocity dependence 
on the scale anomaly of the field. 
We show that there is not a deep interconnection between 
such scale anomaly and the phenomenon of birefringence.  
Finally, we end with some comments concerning the new effects 
that are to be expected for such nonlinear photons.  
Typical Riemannian calculations are presented in appendix \ref{Ageo}, 
in which it is shown that the standard geodesic assumption 
for photons also holds for the nonlinear case.  
Connections with Rainich-Wheeler formalism \cite{Wheeler2} 
are also provided in appendix \ref{rainich}.

\subsection{Definitions and notations}

We call the electromagnetic tensor $F_{\mu\nu}$, while its dual
$\Fstar_{\mu\nu}$ is 
\begin{equation}
\Fstar_{\alpha\beta} \doteq 
\frac{1}{2}\eta_{\alpha\beta}\mbox{}^{\mu\nu}F_{\mu\nu},
\label{0001}
\end{equation}
where $\eta_{\alpha\beta\mu\nu}$ is the completely antisymmetric 
Levi-Civita tensor; the Minkowski metric tensor is represented 
by its standard form $\eta^{\mu\nu}.$ 
The two invariants constructed with these tensors are defined as 
\begin{equation}
F \doteq F^{\mu\nu} \,F_{\mu\nu},
\end{equation}
\begin{equation}
G \doteq F^{\mu\nu} \,\Fstar_{\mu\nu}.
\end{equation}

Once the modifications of the vacuum which will be dealt here 
do not break the gauge invariance of the theory, the 
general form of the modified Lagrangian for electrodynamics 
may be written as a functional of the above 
invariants, that is, 
$$L = L(F,\,G).$$
We denote by $L_F$ and $L_G$ the derivatives of the Lagrangian $L$ 
with respect to the invariant $F$ and $G$, respectively; 
and similarly for the higher order derivatives.  
We are particularly interested in the derivation 
of the characteristic surfaces which guide the propagation 
of the field discontinuities.  

Let $\Sigma$ 
be a surface of discontinuity for the electromagnetic field.  
Following Hadamard \cite{Hadamard} we assume that 
the field itself is continuous when crossing $\Sigma$, 
while its first derivative presents a finite discontinuity.  
We accordingly set
\begin{equation}
\label{[F]}
[F_{\mu\nu}]_{\Sigma} = 0,
\label{gw1}
\end{equation} 
and   
\begin{equation}
\label{[Fnu]}
[\partial_{\lambda}\,F_{\mu\nu}]_{\Sigma} = f_{\mu\nu} k_{\lambda},
\label{gw2}
\end{equation} 
in which the symbol 
$$[J]_\Sigma \equiv \lim_{\delta\rightarrow 0^+} 
\left(J|_{\Sigma+\delta}-J|_{\Sigma-\delta}\right)$$ 
represents the discontinuity of the arbitrary function $J$ 
through the surface $\Sigma$ 
characterized by the equation $\Sigma(x^{\mu}) = constant$.  
The tensor $f_{\mu\nu}$ is called the discontinuity of the field, and 
\begin{equation}
k_\lambda=\partial_{\lambda}\,\Sigma
\label{000111}
\end{equation}
is the propagation vector.  

\section{The method of the effective geometry}
\label{geometry}

\subsection{One-parameter Lagrangians}
Our generic purpose in this paper is to investigate the effects of 
nonlinearities in the equation of evolution of electromagnetic 
waves.  We will restrict the analysis in this section to the 
simple class of gauge invariant Lagrangians defined by 
\begin{equation}
\label{L(F)}
L=L(F).
\end{equation}
From the least action principle we obtain the field equation 
\begin{equation}
\partial_{\mu}\,\left(L_{F}F^{\mu\nu}\right) = 0.
\label{previous}
\end{equation}

Applying conditions (\ref{[F]}) and (\ref{[Fnu]}) for 
the discontinuity of the field equation (\ref{previous}) 
through $\Sigma$ we obtain
\begin{equation}
L_{F}f^{\mu\nu}\, k_{\nu} + 2 L_{FF} \,\xi F^{\mu\nu} k_{\nu} = 0,
\label{gw3}
\end{equation} 
where $\xi$ is defined by 
\begin{equation}
\xi  \doteq  F^{\alpha\beta} \, f_{\alpha\beta}.
\end{equation}
The consequence of such discontinuity in the cyclic identity is
\begin{equation}
f_{\mu\nu} k_{\lambda} + f_{\nu\lambda} k_{\mu} + 
f_{\lambda\mu} k_{\nu} = 0.
\label{gw4}
\end{equation}
In order to obtain a scalar relation we contract this 
equation with $k_{\alpha}\eta^{\alpha\lambda}F^{\mu\nu}$, resulting 
\begin{equation}
\label{gw33}
\xi k_\nu k_{\mu}\eta^{\mu\nu} + 2F^{\mu\nu} 
f_\nu{}^\lambda k_\lambda k_\mu = 0. 
\end{equation}
Let us consider the case in which  $\xi$ does not vanish%
\footnote{For the case in which $\xi = 0$, the quantity $f_{\mu\nu}$ 
is a singular two-form.  Following Lichnerowicz \cite{Lichne}, 
it can be decomposed in terms of the propagating vector $k_{\mu}$ 
and a spacelike vector $a_\mu=a\,\epsilon_\mu$ orthogonal to $k_\mu$, 
in which $\epsilon_{\mu}$ is the normalized polarization vector. 
Hence, we can write $f_{\mu\nu}=k_\mu\,a_\nu- k_\nu\,a_\mu$ on $\Sigma$.  
From equation (\ref{gw3}) it follows that $f^{\mu\nu}k_\nu=0$, 
and contracting (\ref{gw4}) with $\eta^{\lambda\rho}k_\rho$ 
yields $f_{\mu\nu}\eta^{\alpha\beta}k_\alpha k_\beta=0$.  
Therefore, such modes propagate 
along standard null geodesics in Minkowski spacetime.  }.  
From equations (\ref{gw3}) and (\ref{gw33}) 
we obtain the propagation equation for the field discontinuities
as given by
\begin{equation}
\left(L_F\eta^{\mu\nu} - 4L_{FF}F^{\mu\alpha}F_\alpha{}^\nu\right) 
k_\mu k_\nu = 0.
\label{gww4}
\end{equation}

Expression (\ref{gww4}) suggests that one can interpret the 
self-interaction of the background field $F^{\mu\nu},$ in what concerns 
the propagation of electromagnetic discontinuities (\ref{[Fnu]}), 
as if it had induced a modification on the spacetime metric 
$\eta_{\mu\nu}$, leading to the effective geometry

\begin{equation}
g^{\mu\nu}_{\rm eff} = L_{F}\,\eta^{\mu\nu}  - 
4\, L_{FF} \,{F^{\mu}}_{\alpha} \,F^{\alpha\nu}.
\label{geffec}
\end{equation}
A simple inspection of this equation shows that only in 
the particular case of linear Maxwell electrodynamics 
the discontinuity of the  electromagnetic field 
propagates along null paths in the Minkowski background.

The general expression of the effective geometry can be
equivalently written in terms of the vacuum expectation value (VEV) 
of the energy-momentum tensor, given by 
\begin{equation}
T_{\mu\nu} \equiv \frac{2}{\sqrt{-\gamma}} 
\,\frac{\delta\,\Gamma}{\delta\,\gamma^{\mu\nu}},
\end{equation}
where $\Gamma$ is the effective action
\begin{equation}
\Gamma \doteq \int \,d^{4}x \sqrt{-\gamma}\,L,
\end{equation}
and $\gamma_{\mu\nu}$ is the Minkowski metric written in an arbitrary 
coordinate system; $\gamma$ is the corresponding determinant. 
In the case of one-parameter Lagrangians, $L=L(F),$ we obtain
\begin{equation}
T_{\mu\nu} = - 4 L_{F}\, {F_{\mu}}^{\alpha} \, F_{\alpha\nu} - 
L\,\eta_{\mu\nu},  
\end{equation}
where we have chosen an Euclidean coordinate system in 
which $\gamma_{\mu\nu}$ reduces to $\eta_{\mu\nu}.$ 
In terms of this tensor the effective geometry (\ref{geffec}) 
can be re-written as%
\footnote{For simplicity, we will denote the effective metric 
as $g^{\mu\nu}$ instead of $g^{\mu\nu}_{\rm eff}$ from now on.} 
\begin{equation}
\label{gT}
g^{\mu\nu}=\left(L_F+\frac{L\,L_{FF}}{L_F}\right)\eta^{\mu\nu} 
+\frac{L_{FF}}{L_F}T^{\mu\nu}.
\end{equation}
We remark that, once the modified geometry along which the photon 
propagates depends upon the energy-momentum tensor distribution of 
the background electromagnetic field, 
it is tempting to search for an analogy with the corresponding behavior 
of photons in a gravitational field%
\footnote{Let us emphasize that it is no more than a simple analogy.}.  
We will return to this question in Section \ref{consequences}.  

Therefore, the field discontinuities propagate along null geodesics%
\footnote{The proof that such curve is in fact a geodesic line 
is given in Appendix \ref{Ageo}.} 
in an effective geometry which depends on the field energy distribution.  
Let us point out that, as it is explicitly shown from the above equation, 
the stress-energy distribution of the field is the true responsible for 
the deviation of the geometry, as felt by photons, 
from its Minkowskian form\footnote{For $T_{\mu\nu} = 0$, the conformal 
modification in (\ref{gT}) clearly leaves the photon paths unchanged.}.

In order to show (see Appendix \ref{Ageo}) that the photon path 
is actually a geodesic curve, it is necessary to know 
the inverse $g^{\mu\nu}$ of the effective metric $g_{\nu\lambda}$, defined by 
\begin{equation}
\label{gT11}
g^{\mu\nu}\,g_{\nu\lambda} = \delta^{\mu}_{\lambda}.
\end{equation}
This calculation is simplified if we take into account the 
well known properties: 
\begin{equation}
\Fstar_{\mu\nu} \,F^{\nu\lambda} = -\, \frac{1}{4} \,G\, 
\delta^\lambda_\mu,
\end{equation}
and
\begin{equation}
\Fstar_{\mu\lambda} \,\Fstar^{\lambda\nu} - 
F_{\mu\lambda} \,F^{\lambda\nu} 
= \frac{1}{2} \, F \, \delta^{\nu}_{\mu}.
\end{equation}

Thus the covariant form of the metric can be written in the form:
\begin{equation}
\label{gT22}
g_{\mu\nu}= a\,\eta_{\mu\nu} + b\,T_{\mu\nu},
\end{equation}
in which $a$ and $b$ are given in terms of the Lagrangian and 
its corresponding derivatives by:
\begin{equation}
a = -\,b \,\left(\frac{L_{F}^{2}}{L_{FF}}+L+\frac{1}{2}\,T \right),
\end{equation}
and 
\begin{equation}
b  =  16\, \frac{L_{FF}}{L_{F}}\,\left[ \left(F^{2} + G^{2}\right)\,
L_{FF}^{2} - 16\,\left(L_{F} + F\,L_{FF}\right)^{2}\,\right]^{-\,1},
\end{equation}
where $T = T^{\alpha}_{\alpha}$ 
is the trace of the energy-momentum tensor.

\subsection{Two parameter Lagrangians}

In this section we will go one step further and deal with the general case  
in which the effective action depends upon both invariants, that is
\begin{equation}
\label{L(FG)}
L=L(F,\,G).
\end{equation}
The equations of motion are 
\begin{equation}
\label{field}
\partial_{\nu}\,\left(L_{F}F^{\mu\nu} + 
L_{G}\Fstar^{\mu\nu}\right) = 0.
\end{equation}
Our aim is to examine the propagation 
of the discontinuities in such case.  
Following the same procedure as presented 
in the previous section one gets 
\begin{equation}
[L_F\,f^{\mu\nu}+2A\,F^{\mu\nu}
+2B\,\Fstar^{\mu\nu}]\,k_\nu = 0,
\label{ref1}
\end{equation}
and contracting this expression with $F^\alpha{}_\mu k_\alpha$ and 
with $\Fstar^{\alpha}{}_\mu k_\alpha$, respectively, yields
\begin{equation}
\left[ \xi\,L_{F} + \frac{1}{2}\,B\,G \right]\,\eta^{\mu\nu}\,
k_{\mu}\,k_{\nu} - 2 A  {F^{\nu}}_{\alpha}\, F^{\alpha\mu} k_\nu k_\mu 
= 0
\label{gw331}
\end{equation}
and
\begin{equation}
\left[ \zeta\,L_{F} - B\,F + \frac{1}{2}\, A\,G \right]\,\eta^{\mu\nu}\,
k_{\mu}\,k_{\nu} - 2 B  {F^{\nu}}_{\alpha}\, F^{\alpha\mu} k_\nu k_\mu
= 0.
\label{gw332}
\end{equation}
In these expressions we have set
$$A \doteq 2\,(\xi\,L_{FF} + \zeta\,L_{FG}),$$ 
$$B \doteq 2\,(\xi\,L_{FG} + \zeta\,L_{GG}),$$ 
and $\zeta$ is defined by 
\begin{equation}
\zeta  \doteq  \;\;\Fstar^{\alpha\beta} f_{\alpha\beta}.
\end{equation}
In order to simplify our equations it is worth defining 
the quantity $\Omega \doteq \zeta/\xi$. 
From equations (\ref{gw331}) and (\ref{gw332}) it follows 
\begin{equation}
\Omega^{2}\, \Omega_{1} + \Omega\,\Omega_{2} + \Omega_{3} = 0, 
\label{r22}
\end{equation}
with the quantities $\Omega_i,\,\, i=1,2,3$ given by
\begin{eqnarray}
\Omega_1 & = & - L_FL_{FG}+2FL_{FG} L_{GG} \nonumber\\
\label{Omega1} && + G(L_{GG}^2 - L_{FG}^2),\\
\Omega_2 & = &
(L_F+2GL_{FG})(L_{GG} - L_{FF})  \nonumber\\ 
&& +  2F(L_{FF}L_{GG} + L_{FG}^2),\\
\label{Omega2}
\Omega_3 & = & L_FL_{FG}+2FL_{FF}L_{FG} \nonumber\\
&& + G(L_{FG}^2 - L_{FF}^2).
\label{Omega3}
\end{eqnarray}
The quantity $\Omega$ has two solutions and is given by the algebraic
expression 
\begin{equation}
\Omega_{\scriptscriptstyle \pm} 
= \frac{-\Omega_2 \,\pm\, \sqrt{\Delta}}{2\Omega_1} ,
\label{Omega}
\end{equation}
where 
$$
\Delta \doteq (\Omega_2)^2 - 4\Omega_1\Omega_3.
$$ 
Thus, in the general case we are concerned here, the 
photon paths are kinematically described by 
\begin{equation}
\label{g2ef}
g_{\scriptscriptstyle \pm}^{\mu\nu}\,k_{\mu}\,k_{\nu} = 0,
\end{equation}
where the effective metrics $g_{\scriptscriptstyle \pm}^{\mu\nu}$ 
are given by 
\begin{eqnarray}
g_{\scriptscriptstyle \pm}^{\mu\nu} &=&L_F\eta^{\mu\nu} - 4\left[\left(
L_{FF} + \Omega_{\scriptscriptstyle \pm} L_{FG}\right)F^{\mu}\mbox{}_{\lambda}
F^{\lambda\nu} \right.\nonumber\\
&& \left.+ \left(L_{FG} + \Omega_{\scriptscriptstyle \pm} L_{GG}\right)
F^{\mu}\mbox{}_{\lambda}\Fstar^{\lambda\nu}\right].
\label{geral}
\end{eqnarray}
When the Lagrangian does not depend on the invariant $G$, 
expression (\ref{geral}) reduces to the form (\ref{geffec}).  

As we see, from equation (\ref{Omega}), there will be two possible solutions
for the paths of light, which are related to its different modes of
polarization, indicating that birrefringence phenomena could be described
here, in a general way, depending on the particular theory we shall consider.

The tensor of discontinuities, given by equation (\ref{gw2}), can be
decomposed as
\begin{equation}
f_{\alpha\beta} = a\left(p_{\alpha}k_{\beta} - 
p_{\beta}k_{\alpha}\right)
\end{equation}
where $a$ in the strength of the wavelet and $p_\alpha$ represents
the polarization vector, which is ortogonal to $k_\mu$ and normalized
to unity:
\begin{eqnarray}
p^\alpha k_\alpha &=& 0
\\
p^\alpha p_\alpha &=& -1.
\end{eqnarray}
For a given wave vector $k_\mu$ there will be two linearly independent
polarization vectors $p_\mu$, which satisfy the above conditions.
Introducing $f_{\alpha\beta}$ in the field equations (\ref{ref1}) we obtain
the expression that describes the states of polarization, that is,
\begin{eqnarray}
k^2 p^\mu = &-&\frac{4}{L_F}
\left[L_{FF}F^{\mu\alpha}F^{\nu\beta} + 
L_{GG}\Fstar^{\mu\alpha}\Fstar^{\nu\beta} \right.
\nonumber \\
&+& \left. L_{FG}\left(F^{\mu\alpha}\Fstar^{\nu\beta} +
\Fstar^{\mu\alpha}F^{\nu\beta}\right) 
\right]
k_{\alpha}k_{\beta}p_{\nu}.
\end{eqnarray}
This equation must be solved using each solution of the wave vector 
coming from equation (\ref{g2ef}) when the effective
metrics $g^{\mu\nu}_{\scriptscriptstyle \pm}$ are used. 

From the general expression of the energy-momentum tensor 
for an electromagnetic theory $L=L(F,\,G)$ we have
\begin{equation}
T_{\mu\nu} = - 4 L_{F}\, {F_{\mu}}^{\alpha} \, F_{\alpha\nu} - 
\left( L - G \, L_{G} \right) \,\eta_{\mu\nu}. 
\end{equation}
The scale anomaly is given by the trace 
\begin{equation}
\label{scaleA}
T = 4\, \left( -L + F \,L_{F}  + G \,L_{G} \right).
\end{equation}
We can then re-write the effective geometry in a more 
appealing form in terms of the energy momentum tensor, that is,
\begin{equation}
g^{\mu\nu} = {\cal M}_{\scriptscriptstyle \pm}\,\eta^{\mu\nu} + 
{\cal N}_{\scriptscriptstyle \pm} \,T^{\mu\nu},
\label{AB}
\end{equation}
where the functions $ {\cal M}_{\scriptscriptstyle \pm}$ and 
${\cal N}_{\scriptscriptstyle \pm}$ are given by
\begin{eqnarray}
\label{A}
{\cal M}_{\scriptscriptstyle \pm} & = & L_F + G
\left( L_{FG} + \Omega_{\scriptscriptstyle \pm} L_{GG} \right) +
\nonumber \\
&&+ \frac{1} { L_{F}} \left( L_{FF} + \Omega_{\scriptscriptstyle \pm} 
L_{FG} \right) \left( L -
G L_{G} \right),
\\
\label{B}
{\cal N}_{\scriptscriptstyle \pm} & = & \frac{ 1 }{ L_{F} } \left( L_{FF} 
+ \Omega_{\scriptscriptstyle \pm} L_{FG} \right).
\end{eqnarray}
As a consequence of this, the 
Minkowskian norm of the propagation vector $k_\mu$ reads 
\begin{equation}
\eta^{\mu\nu} k_{\mu}\,k_{\nu} 
= - \frac{{\cal N}_{\scriptscriptstyle \pm}}
{{\cal M}_{\scriptscriptstyle \pm}}
T^{\mu\nu}k_{\mu}k_{\nu}.
\label{labell}
\end{equation}

\subsection{Exceptional Lagrangians}

It seems worth noting that equation (\ref{geral}) contains a remarkable 
result: the velocities of the photon are, in general, doubled. 
There are some exceptional cases, however, for  which the uniqueness 
of the path is guaranteed by the equations of motion \cite{boillat,birula}.  
Such uniqueness occurs for those dynamics described by Lagrangian $L$ 
that satisfy the condition 
$$\Delta = 0.$$ 

The most known example of such uniqueness for the photon velocity 
in a nonlinear theory is the Born-Infeld electrodynamics. 
Let us pause for a while in order to make the following remark.  
In the case of the Born-Infeld theory 
all quantities $\Omega_i,\,\, i=1,2,3$ vanish identically.
Hence, in this situation we cannot obtain the effective geometry from 
equation (\ref{geral}). In this very exceptional case we proceed as
follows. Let us return to the original equation (\ref{gw331}).  
Now, the Lagrangian for the Born-Infeld model 
is provided by the expression 
\begin{equation}
L = \sqrt{b^{4} + \frac{1}{2}\,b^{2}\,F - \frac{1}{16}\,G^2} - b^{2}. 
\label{born}
\end{equation}
Substituting this form of $L$ into  equation (\ref{gw331}) 
we obtain the unique characteristic equation
\begin{equation}
A\left[\left(b^{2} + \frac{1}{2}\,F\right)\, \eta^{\mu\nu} 
+ F^{\mu}\mbox{}_{\lambda}F^{\lambda\nu}\right] k_{\mu}\,k_{\nu} = 0,
\label{quem1}
\end{equation}
thus yielding (for $A \neq 0$)
\begin{equation}
g^{\mu\nu} = (b^{2} + \frac{1}{2}\,F)\,\eta^{\mu\nu} 
+ F^{\mu}\mbox{}_{\lambda}F^{\lambda\nu},
\label{quem2}
\end{equation}
which does not show birefringence.  This formula 
was obtained for the first time by Plebansky \cite{Plebansky}.  

\section{Some remarkable consequences}
\label{consequences}
\subsection{Electromagnetic traps}
\label{traps}

The possibility of writing the characteristic equation for 
nonlinear electrodynamics in terms of an effective modification 
of the spacetime metric has some unexpected and wide-ranging 
consequences. One of these concerns to the existence of 
trapped photons in a compact domain. 
Such a configuration is made possible due to the 
nonlinearity of electrodynamics, and 
has a striking resemblance to gravitational black holes, 
although not presenting all properties of the latter. 

We will concentrate here on a toy model, just to exhibit the 
possibility of new phenomena induced by the nonlinearities.
The well known solution $\eta^{\mu\nu}k_{\mu}k_{\nu} = 0$, that
also appears in this case, will not be considered. 
Let us start with a static and spherically symmetric 
field for the case $L = L(F).$ The source is an electric 
monopole located at the origin of the spherical coordinate system 
$(t,\,r,\,\theta,\,\varphi)$. 
We set for the non-zero components of the electromagnetic field 
the form $$F_{tr} = E(r).$$
The equation of motion are easily solved, yielding
\begin{equation}
L_{F}\,E = \frac{Q}{r^{2}}.
\label{ee1}
\end{equation}
The corresponding effective geometry (\ref{geffec}) is given by
\begin{equation}
g^{tt} = -\,g^{rr} = L_{F} - 4\,L_{FF}\,E^{2},
\label{ee2}
\end{equation}
while the remaining non-zero components have values proportional 
to those of Minkowski geometry, 
\begin{eqnarray}
g^{\theta\theta} &=& -\frac{1}{r^{2}}L_F,
\label{ee3}\\
g^{\varphi\varphi} &=& -\frac{1}{r^{2}\,\sin^{2}\theta} L_F.
\label{ee4}
\end{eqnarray}
From equation (\ref{ee2}) it follows that it is possible to 
envisage the existence of a region ${\cal D}$  defined by 
some finite radius $r = r_c$ such that $g^{rr}(r_c)$ vanishes. 
The metric component $g^{tt}$ also vanishes at ${\cal D}$. 
Then, coordinates $r$ and $t$ 
interchange their roles when crossing ${\cal D},$ that is 
$g^{tt}(r > r_c) > 0$, $g^{rr}(r > r_c) < 0$, and 
$g^{tt}(r < r_c) < 0$, $g^{rr}(r < r_c) > 0$.  
Let us note, however, that the existence of such $r_c$ 
implies that there is a further undesirable 
consequence concerning the value of the 
electric field. Indeed, equation (\ref{ee1}) yields
\begin{equation}
r \frac{\partial r}{\partial E} = 
\frac{Q}{2\, L_{F}^{2}\,E^{2}}\, g^{rr}.
\label{ee5}
\end{equation}
Thus it follows from this that, at ${\cal D}$, there is a possibility
of the existence of a field barrier, that is a limitation of the domain 
of its existence due to the possibility of the inverse function 
$r = r(E)$ to have an extremum. In order to avoid such limitation 
of the domain of definition for the electric field, the 
theory must be such that the second derivative vanishes at 
$r_c$, that is 
$$\frac{\partial\,g^{rr}}{\partial\,E}|_{\cal D} = 0.$$ 
Besides, one must impose the supplementary condition 
$$\frac{\partial^{2}\,g^{rr}}{\partial\,E^{2}}|_{\cal D} \neq 0.$$
Thus, if the theory $L = L(F)$ allows this kind of solution, 
then three typical properties of a trapped region appear:
\begin{itemize}
\item{There exists a null surface ${\cal D},$ defined by $r = r_c$ 
in the effective geometry.}
\item{Coordinates $t$ and $r$ interchange their role 
when crossing ${\cal D}$.}
\item{Light cones inside the region bounded by ${\cal D}$ 
are directed towards the origin of the $r$-coordinate which plays the role, 
in this domain and only for photon propagation, of a time-like coordinate.}
\end{itemize}

These theories should be further examined, since from what we have 
seen above, they may be the germ of the existence of the 
electromagnetic version of the gravitational black hole.  
Let us now turn our examination 
to a very similar situation inside material media. 

\subsection{Wave propagation in nonlinear dielectric media}

It is possible to describe the wave propagation, governed by Maxwell 
electrodynamics, inside a dielectric, in terms of a modification of the 
underlying spacetime geometry using the framework developed above.
The electromagnetic field is represented by two antisymmetric
tensors, the electromagnetic field $F_{\mu\nu}$ and 
the polarization $P_{\mu\nu}$. These tensors are decomposed, 
in the standard way, into their corresponding electric and magnetic
parts as seen by an observer which moves with velocity $v_\mu$.  
We can write:
\begin{equation}
F_{\mu\nu} = E_{\mu}\, v_{\nu} - E_{\nu}\, v_{\mu} + 
\eta^{\rho\sigma}\mbox{}_{\mu\nu} \, v_{\rho} \, H_{\sigma},
\end{equation}
\begin{equation}
P_{\mu\nu} = D_{\mu}\, v_{\nu} - D_{\nu}\, v_{\mu} + 
\eta^{\rho\sigma}\mbox{}_{\mu\nu} \, v_{\rho} \, B_{\sigma}.
\end{equation}

Following Hadamard, we consider the discontinuities on the fields 
as given by
\begin{eqnarray}
\left[\bigtriangledown_{\lambda}\, E_{\mu}\right]_{\Sigma} 
&=& k_{\lambda}\, e_{\mu},
\nonumber\\
\left[\bigtriangledown_{\lambda}\, D_{\mu}\right]_{\Sigma} 
&=& k_{\lambda}\, d_{\mu},
\nonumber\\
\left[\bigtriangledown_{\lambda}\, H_{\mu}\right]_{\Sigma} 
&=& k_{\lambda}\, h_{\mu},
\nonumber\\
\left[\bigtriangledown_{\lambda}\, B_{\mu}\right]_{\Sigma} 
&=& k_{\lambda}\, b_{\mu}.
\end{eqnarray} 

For the simplest linear case, in which we have
\begin{eqnarray}
D_{\alpha} &=&  \epsilon \, E_{\alpha} ,
\\
B_{\alpha} &=&   \frac{H_{\alpha}}{\mu} ,
\end{eqnarray}
it follows that 
\begin{eqnarray}
d_{\alpha} &=&  \epsilon \, e_{\alpha} ,
\\
b_{\alpha} &=&   \frac{h_{\alpha}}{\mu}. 
\end{eqnarray}
After a straightforward calculation one obtains 
\begin{equation}
k_{\mu}\, k_{\nu} \left[ \gamma^{\mu\nu} + (\epsilon\, \mu - 1) v^{\mu}
\,v^{\nu} \right] = 0.
\end{equation}
Let us generalize this situation for the nonlinear case. 
Maxwell equations are given by
\begin{equation}
\partial^{\nu} \,\Fstar_{\mu\nu} = 0,
\end{equation}
\begin{equation}
\partial^{\nu} \,P_{\mu\nu} = 0.
\label{polar}
\end{equation}
For electrostatic fields inside 
isotropic dielectrics it follows that $P^{\mu\nu}$ and 
$F^{\mu\nu}$ are related by 
\begin{equation}
\label{rel}
P_{\mu\nu} = \epsilon(E) F_{\mu\nu}.
\end{equation}
where $\epsilon$ is the electric susceptibility. 
In the general case, for $\epsilon = 
\epsilon\,(E)$ we simplify our calculation if we note that 
we can relate the equation of wave propagation to the previous analysis 
on vacuum polarization (\ref{L(F)}) by means of the identification 
\begin{equation}
L_{F}  \longrightarrow \epsilon,
\end{equation}
which implies 
\begin{equation}
L_{FF} \longrightarrow -\,\frac{\epsilon\rq}{4\,E},
\end{equation}
in which $\epsilon'\equiv{d\,\epsilon}/{d\,E}$.  
Therefore, the simple class of effective Lagrangians (\ref{L(F)}) 
may be used as a convenient description of Maxwell theory 
inside isotropic nonlinear dielectric media; 
conversely, results obtained in the latter context 
can as well be similarly restated in the former one.  

In a nonlinear dielectric medium the polarization induced by 
an external electric field is described by expressing 
the scalar function $\epsilon$ as a power series in terms of 
the field strength $E$: 
\begin{equation}
\label{xxx7}
\epsilon = \chi_{1} + \chi_{2}\, E +  \chi_{3}\, E^2 + 
\chi_{4}\, E^3 + \ldots
\end{equation}
where the constants $\chi_n$ are known as 
the n-order nonlinear optical susceptibility.  
Note that we are using the standard convention \cite{Plebansky} 
which relates $\chi_n$ with the expansion of the polarization vector.  
For this case the effective geometry is given by 
\begin{equation}
\label{gef}
g^{\mu\nu} = \epsilon\, \eta^{\mu\nu} + \frac{\epsilon'}{E}F^{\mu}
{}_{\alpha}F^{\alpha\nu}.
\end{equation}
It can also be re-written in the form
\begin{equation}
\label{gef22}
g^{\mu\nu} = \epsilon\, \eta^{\mu\nu} - \frac{\epsilon\rq}{E} \left( 
E^{\mu}\,E^{\nu} - E^2\, \delta^{\mu}_t \,\delta^{\nu}_t \right),
\end{equation}
where $E^2 \equiv -\,E_{\alpha}\,E^{\alpha} > \,0.$
In other words,
\begin{eqnarray}
\label{gef223}
g^{tt} &=& \epsilon  + \epsilon' E \\
\label{gef224}
g^{ij} &=& -\,\epsilon\,\delta^{ij} - \frac{\epsilon'}{E}\, E^i\,E^j .
\end{eqnarray}
This shows that the discontinuities of the electromagnetic field 
inside a nonlinear dielectric medium propagates along null cones 
of an effective geometry which depends on the characteristics 
of the medium given by equation (\ref{gef}).  It seems worth 
investigating under what conditions of the $\epsilon$ dependence on 
$E$ a kind of horizon barrier should appear for the photon inside 
a dielectric. We will return to this problem elsewhere.

\subsection{Photon path}

From what we have learned above it follows that the equation of 
motion of the photon in a nonlinear regime is given by the variational 
principle
\begin{equation}
\delta \,\int\, ds  = 0,
\label{2121}
\end{equation}
in which the fundamental length is constructed with 
the effective metric, $ds^{2} = g_{\mu\nu}\,dx^{\mu}\,dx^{\nu}$.  
For the particular case which was examined in the previous section 
concerning the static and spherically symmetric configuration 
it becomes
\begin{equation}
\delta \,\int \left(g_{tt} \dot{t}^{2} + g_{rr} \dot{r}^{2} + 
g_{\theta\theta} \dot{\theta}^{2} + 
g_{\varphi\varphi} \dot{\varphi}^{2}\right) ds \, = 0,
\label{21212}
\end{equation}
in which a dot means derivative with respect to the 
fundamental lenght variable $s.$
The equation for the angular variable $\theta$ shows that we can choose 
conveniently the initial condition such that $\theta$ remains constant. 
From an analogy with the planetary motion we set $\theta = {\pi}/{2}$.  
The corresponding equations of the remaining variables are 
\begin{equation}
r^{2} \dot{\varphi} =h_o,
\label{geo1}
\end{equation}
\begin{equation}
g_{tt}\,\dot{t} =E_o,
\label{geo2}
\end{equation}
where $h_o$ and $E_o$ are constants of motion.  
It is rather convenient to obtain the equation for $r$ by making 
use of the fact that we are dealing with a null curve, and set
\begin{equation}
g_{tt}\,\dot{t}^{2} + g_{rr}\,\dot{r}^{2} 
+ g_{\varphi\varphi}\,\dot{\varphi}^{2} = 0.
\label{geo3}
\end{equation}
Thus, using the above equations for the evolution of $t$, $\theta$ 
and $\varphi$ we obtain:
\begin{equation}
\dot{r}^{2} = E_o^{2} - V(r),
\label{geo4}
\end{equation}
in which the potential  $V(r)$ takes the form
\begin{equation}
V(r) = \frac{-\,E_o^2}{g_{tt}^2} + 
 \frac{h_o^2}{g_{tt}\,r^{2}} + E_o^2.
\label{geo5}
\end{equation}

\subsubsection{Circular orbits}

The above set of equations allows the possibility of the 
existence of circular orbits $r = r_o =constant$ for the photon.  
In this case, it is sufficient for the value of the $tt-$component 
for the effective metric at the point $r_o$ to take the value
\begin{equation}
g_{tt}(r_o) = \frac{1}{l^2} \,r_o^2,
\end{equation}
in which we have, for comparison with the planetary motion, defined
the impact parameter $l \doteq{h_o}/{E_o}$.  
It is a rather simple and straightforward matter to show that 
such orbits are unstable, as one should suspect.

\section{Non trivial vacua}
\label{vacua}

There have been some comments in the literature regarding a 
possible connection between the change of light velocity 
in modified vacua and the scale anomaly.  The first one 
to speculate upon such a correlation was Shore \cite{Shore}, and 
Dittrich and Gies \cite{Dittrich} disproved, using the 
sum over polarization states method, the existence of
such correlation for the case of a general Lagrangian $L(F,G)$. 

Using the method of the effective geometry we are able to
show that there is no deep correlation between the 
existence of the velocity shifts induced by the nonlinear 
quantum fluctuation and the scale anomaly. 
We will consider this problem in a general framework.  
In other words, let us  analyze the photon velocity 
in an arbitrary case, without specifying a particular theory.
From what we have learned above one could expect that the regime 
of nonlinearity induced by the modification of the electrodynamic 
vacuum to be the sufficient condition to modify the photon 
velocity.  However, this is not true in general, as 
it can be seen below. 

\subsection{Conformal vacuum}

An important example of modification of the action occurs for the case 
in which the quantity ${\cal N}_{\scriptscriptstyle \pm}$ vanishes.  In this case the net effect 
of this action on the photon velocity 
is indistinct from the one produced by the classical vacuum.  
Could this occur for the case in which $T \neq 0 $?  
This is answered by the following lemma.
\begin{lemma}
\label{lemmaA}
There exist nonlinear modifications of electrodynamics 
which present a non identically null anomaly such that 
the field discontinuities propagate along Minkowskian paths.  
\end{lemma}
The proof is immediate. From equation (\ref{AB}), 
the general condition for such statement to hold 
is expressed by
\begin{equation}
L_{FF} + \Omega_{\scriptscriptstyle \pm} \, L_{F\,G} = 0.
\label{Bzero}
\end{equation}
Any nonlinear action which satisfies this 
condition is such that the photon propagation occurs in an effective 
geometry $g^{\mu\nu}$, which is conformal to the Minkowskian one.  
In this theory, the photon presents the same light-cone structure 
as it does in the linear Maxwell electrodynamics.  
The important point to consider here is that this situation 
occurs even in the presence of a non-vanishing scale anomaly. 

Inserting $\Omega_{\scriptscriptstyle \pm}
= -L_{FF}/L_{FG}$ from (\ref{Bzero}) 
into equation (\ref{r22}) we obtain, 
after some algebra, the condition 
\begin{equation}
\left(L_{FF}L_{GG}-L_{FG}^2\right)\left[L_F L_{FG}
+G\left(L_{FF}L_{GG}-L_{FG}^2\right)\right] = 0.
\label{ref2}
\end{equation}
It remains to show that the spectrum of common solutions 
of (\ref{Bzero}) and (\ref{ref2}) does not imply $T=0$.  
This can be explicitly shown 
for the particular class of nonlinear Lagrangians given by 
\begin{equation}
L = -\frac{1}{4}F  + f(G),
\label{ref3}
\end{equation}
for which $L_{FF}=0$ and $L_{FG}=0$.  
The scale anomaly (\ref{scaleA}) 
for the class shown in (\ref{ref3}) takes the form
\begin{equation}
T = 4\left(G\,\frac{\partial\,f}{\partial\,G} -f\right),
\label{ref4}
\end{equation}
which vanishes only if $f(G)$ is a linear function of $G$.  
Expression (\ref{Omega}) yields $\Omega_{\scriptscriptstyle +}=0$ 
and $\Omega_{\scriptscriptstyle -}=1/(4G\,\partial^2f/\partial\,G^2)$, 
from which (\ref{A}) gives ${\cal M}_{\scriptscriptstyle \pm}
=-1/4$ and ${\cal M}_{\scriptscriptstyle \pm}=0$, 
respectively.  The apparent singularity in (\ref{AB}) 
for ${\cal M}_{\scriptscriptstyle \pm}
=0$ can be circumvented by carefully 
returning to the original expression (\ref{ref1}), 
which reduces in this case to 
\begin{equation}
f^{\mu\nu}k_\nu=\frac{2\chi}{G}\Fstar^{\mu\nu}k_\nu.
\label{ref5}
\end{equation}
It follows that $\eta^{\mu\nu}k_\mu k_\nu=0$, thus adequately 
describing both solutions associated with (\ref{ref3}).  

\subsection{Non-anomalous vacuum}

There is no better way to demonstrate the independence of 
the concepts of light velocity modification and the anomaly 
than to consider the converse situation of Lemma \ref{lemmaA}, 
for which there is no anomaly at all.  This is provided by the 
following 
\begin{lemma}
\label{lemmaB}
In the absence of scale anomaly the effective metric is not 
necessarily conformally flat.
\end{lemma}
Indeed, let us set 
\begin{equation}
\label{anomaly}
T = 0.
\end{equation}
From (\ref{scaleA}) it follows
\begin{equation}
\label{ino}
L = F \,L_{F} + G \, L_{G},
\end{equation}
which leads to 
\begin{equation}
F \, L_{FF} + G \, L_{G\,F} = 0,
\end{equation}
and
\begin{equation}
G \, L_{G\,G} + F \, L_{F\, G} = 0.
\end{equation}
In order for the effective metric to be conformally flat 
the Lagrangian should obey, besides the above relations, 
the condition (\ref{Bzero}). Then, it is straightforward to 
show that Lagrangians satisfying all these requirements 
must satisfy the condition 
\begin{equation}
\label{LFF}
L_{FF} = 0.
\end{equation}
Hence, when there is no anomaly, the unique case in which the 
effective geometry coincides with the Minkowski one is the linear 
Maxwell theory.  This property allows us to conclude that 
in the framework of the effective action {\bf there is no need 
for a scale anomaly to induce light velocity shifts}.  

It remains to show that the spectrum of solutions of equation 
(\ref{anomaly}) does not necessarily reduce to the linear case.  
It is interesting to point out that not only one 
but a particular set of nonlinear Lagrangians can be obtained.  
Indeed, it is immediately shown that 
\begin{equation}
L = G\,f(F/G)   
\end{equation}
satisfies equation (\ref{ino}) for arbitrary functions $f(F/G)$.

\subsection{Euler-Heisenberg vacuum}

The effective action for electrodynamics due to one-loop quantum 
corrections was calculated by Heisenberg and Euler \cite{Euler}.  
For the low-frequency limit $\nu \ll m_ec^2/h$ 
the effective Lagrangian takes the form 
\begin{equation}
L = - \frac{1}{4} F + \frac{\mu}{4} 
\left( F^2 + \frac{7}{4} G^2\right),
\protect\label{Euler}
\end{equation}
with 
\begin{equation}
\protect\label{mu}
\mu \doteq \frac{2}{45} \, \alpha^2 \, 
\left(\frac{\hbar}{m_{e}\,c}\right)^3 \, \frac{1}{m_{e}\,c^2},
\end{equation}
where $\alpha$ is the fine-structure constant.

The trace of the corresponding modified energy-momentum tensor reads
\begin{equation}
T = \mu \,\left( F^{2} + \frac{7}{4}\, {G}^2 \right).
\end{equation}
The coefficients ${\cal M}_{\scriptscriptstyle \pm}$ 
and ${\cal N}_{\scriptscriptstyle \pm}$ of the associated 
metric tensor at the first order of approximation in $\mu$ constant
are:
\begin{eqnarray}
{\cal M}_{\scriptscriptstyle -} &=& - \frac{1}{4} +\frac{\mu}{2} F  \\
{\cal M}_{\scriptscriptstyle +} &=& - \frac{1}{7} +\frac{5\mu}{7} F \\
{\cal N}_{\scriptscriptstyle \pm} &=& -  2\mu .
\end{eqnarray}
We note that only in the case where both invariants 
$F$ and $G$ vanish does the trace anomaly disappear.  
In this case the above coefficients become 
${\cal M}_{\scriptscriptstyle -} = - \frac{1}{4}$,
${\cal M}_{\scriptscriptstyle +} = - \frac{1}{7}$ 
and ${\cal N}_{\scriptscriptstyle \pm} = -2\mu$. 

The effective geometries yield 
\begin{eqnarray} 
g^{\mu\nu}_{\scriptscriptstyle -}  &=& \left(-\,\frac{1}{4} + \mu\,F
\right) \,\eta^{\mu\nu} - 2\,\mu\,T^{\mu\nu}
\\
g^{\mu\nu}_{\scriptscriptstyle +}  &=& \left(-\,\frac{1}{7} + \mu\,F
\right) \,\eta^{\mu\nu} - 2\,\mu\,T^{\mu\nu}.
\end{eqnarray}
This implies that there exist two paths of light, one for each
polarization mode. In other words, birrefringence effects are present
in Euler-Heisenberg electromagnetism, as it is well known in literature.
From equation (\ref{labell}) we can write the associated expressions for
the wave vector propagation:
\begin{eqnarray} 
k^2_{-}  &=& -8\,\mu\,T^{\mu\nu}k_\mu k_\nu
\\
k^2_{+}  &=& -14\,\mu\,T^{\mu\nu}k_\mu k_\nu.
\end{eqnarray}
Taking the average over polarization modes we obtain the well known
formula:
\begin{eqnarray} 
k^2  &=& -11\,\mu\,T^{\mu\nu}k_\mu k_\nu.
\end{eqnarray}

\section{Conclusion}

From what we have learned in this paper we can state that the 
propagation of discontinuities of electromagnetic field 
in a nonlinear regime 
(as it occurs, for instance, in dielectrics or in modified QED vacua)
can be described in terms of an effective modification of the 
Minkowskian geometry of spacetime.  Such interpretation 
is an immediate consequence of the analysis we presented here. 
We would like to point out that it is not impossible to envisage 
the case in which the characteristic surfaces 
along which photons propagate, in nonlinear electrodynamics, 
could appear as space-like hypersurfaces in Minkowski spacetime.  
We will return to this question in a forthcoming paper.  

This description also allows us to recognize a striking analogy between 
photon propagation in nonlinear electrodynamics and its behavior in an 
external gravitational field.  For in both cases the geometry 
is modified by a nonlinear process.  It is clear that such analogy can 
not be pushed very far, since in the gravitational case 
the modified geometry is observed by any kind of matter and energy
(including gravitational energy --- at least in 
the general relativity theory%
\footnote{For the recent proposal of NDL theory of gravity \cite{ndl} 
this would not be the case.}) and in the electromagnetic case 
this modified geometry is observed only by the nonlinear photons.
Moreover, we would like to stress that we proved the existence of
Minkowski geodesics at non-vanishing scale anomaly and vice versa.

This analogy certainly deserves further examination, 
since it may provide for the existence of an electromagnetic analogue 
of the gravitational black hole, as was shown. 

\acknowledgements

We would like to thank the referees of the manuscript 
for their useful suggestions, which led us to improve this work.  
V. A. De Lorenci and R. Klippert are grateful to Dr. N. F. Svaiter 
for valuable discussions on non trivial QED vacua.  
This work was partially supported by {\em Conselho Nacional de 
Desenvolvimento Cient\'{\i}fico e Tecnol\'ogico} (CNPq)
and {\em Funda\c{c}\~ao de Amparo \`a 
Pesquisa do Estado do Rio de Janeiro} (FAPERJ) of Brazil.

\appendix
\section{The effective null geodesics}
\label{Ageo}

The geometrical relevance of the effective geometry 
(\ref{geffec}) goes beyond its immediate definition.  
Indeed, as follows it will be shown that 
the integral curves of the vector $k_\nu$ ({\em i.e.}, 
the photons trajectories) are in fact geodesics.  
In order to achieve this result it will be required 
an underlying Riemannian structure for the manifold 
associated with the effective geometry.  In other words 
this implies a set of Levi-Civita connection coefficients 
$\Gamma^\alpha\mbox{}_{\mu\nu}=\Gamma^\alpha\mbox{}_{\nu\mu}$, 
by means of which there exists a covariant differential operator 
$\nabla_\lambda$ (the {\em covariant derivative}) such that 
\begin{equation}
\label{Riemann}
\nabla_\lambda g^{\mu\nu}\equiv 
g^{\mu\nu}\mbox{}_{;\,\lambda} 
\equiv g^{\mu\nu}\mbox{}_{,\,\lambda} + 
\Gamma^\mu\mbox{}_{\sigma\lambda}g^{\sigma\nu} + 
\Gamma^\nu\mbox{}_{\sigma\lambda}g^{\sigma\mu}=0.  
\end{equation}
From (\ref{Riemann}) it follows that 
the effective connection coefficients are completely determined 
from the effective geometry by the usual Christoffel formula.  

Contracting (\ref{Riemann}) with $k_\mu k_\nu$ results 
\begin{equation}
\label{N15}
k_\mu k_\nu g^{\mu\nu}\mbox{}_{,\,\lambda} = 
-2k_\mu k_\nu\Gamma^\mu\mbox{}_{\sigma\lambda}g^{\sigma\nu}.
\end{equation}
Differentiating (\ref{g2ef}) we have 
\begin{equation}
\label{N16}
2k_{\mu,\,\lambda}k_\nu g^{\mu\nu} + 
k_\mu k_\nu g^{\mu\nu}\mbox{}_{,\,\lambda} = 0.
\end{equation}
Inserting (\ref{N15}) 
for the last term on the left hand side of (\ref{N16}) we obtain 
\begin{equation}
\label{N18}
g^{\mu\nu}k_{\mu;\,\lambda}k_\nu \equiv 
g^{\mu\nu}\left(k_{\mu,\,\lambda} - 
\Gamma^\sigma\mbox{}_{\mu\lambda}k_\sigma\right)k_\nu = 0.
\end{equation}
As the propagation vector $k_\mu=\Sigma_{,\,\mu}$ 
is an exact gradient one can write 
$k_{\mu;\,\lambda}=k_{\lambda;\,\mu}$.  
With this identity and defining $k^\mu\doteq g^{\mu\nu}k_\nu$ 
equation (\ref{N18}) reads 
\begin{equation}
\label{geodesic}
k_{\mu;\,\lambda}k^\lambda = 0,
\end{equation}
which states that $k_\mu$ is a geodesic vector.  
By remembering it is also a null vector 
(with respect to the effective geometry $g^{\mu\nu}$), 
it follows that its integral curves are therefore null geodesics.  

\section{Generalization of the Rainich-Wheeler already unified 
program}
\label{rainich}

We show a remarkable property of the class of nonlinear theories 
which are free of anomaly. 
From the definition of $T_{\mu\nu}$ in the case of two parameter 
Lagrangians we have
\begin{equation}
T_{\mu\nu} = - 4 L_{F}\, {F_{\mu}}^{\alpha} \, F_{\alpha\nu} - 
\left( L - G \, L_{G} \right) \,\eta_{\mu\nu}.
\end{equation}
In the linear case it follows that the square 
$T_{\mu\alpha} \,T^{\alpha\nu}$ of this tensor 
is proportional to the identity matrix $\delta_{\mu}^{\nu}$. 
This {\it square} property was used by Rainich and Wheeler%
\footnote{See for instance \cite{Wheeler1,Wheeler2}.} 
to set a basis of an extension of the geometrization 
program beyond gravitational interaction, the so-called 
{\it already unified program}.  
It is important to remark that this property 
no longer holds for a general nonlinear Lagrangian. 
Indeed, a straightforward calculation gives
\begin{equation}
T_{\mu\alpha} \,T^{\alpha\nu} = m\,\delta_{\mu}^{\nu} +
\frac{1}{2}\,T\, T_{\mu}^{\nu},
\label{TW}
\end{equation} 
in which $m$ is given by
\begin{equation}
m = L_{F}^{2}\, (F^{2} + G^{2}) - \frac{T^{2}}{16},
\end{equation}
and the trace $T \doteq T_{\alpha}^{\alpha}$ takes the value 
\begin{equation}
T = 4\,\left( F\,L_{F} + G\,L_{G} - L \right).
\end{equation}
From expression (\ref{TW}) it follows the interesting result that 
the {\it square} property remains valid for the class of theories 
which do not present scale anomaly. 

We can thus state the Rainich-Wheeler conditions
\begin{mathletters}
\label{RW}
\begin{eqnarray}
& R_{\mu\nu}\,R^{\nu\alpha} = \frac{1}{4}\, 
R_{\lambda\eta} R^{\lambda\eta} \,\delta_{\mu}^{\alpha},
\label{RW1}
\\
& R = 0,
\label{RW2}
\end{eqnarray}
where $R\doteq R^\alpha_\alpha$ is the curvature Ricci scalar, 
and the assumption that 
\begin{equation}
\left({R_{\mu\alpha;\beta}\,\eta^{\alpha\beta\rho\nu}
\,R^{\mu}_\rho}\right) \left(R_{\epsilon\lambda}
\,R^{\epsilon\lambda}\right)^{-\,1}
\label{RW3}
\end{equation}
\end{mathletters}
is a gradient.  
It is straightforward to show that property (\ref{RW3}) 
holds also for the nonlinear case.    
$R_{\mu\nu}$ is the curvature Ricci tensor, 
which satisfies Einstein equations 
\begin{equation}
R_{\mu\nu}-\frac{1}{2}\,R g_{\mu\nu} = - \kappa T_{\mu\nu},
\end{equation}
where $\kappa$ is Einstein's gravitational constant.  
Relations (\ref{RW}) hold for a general nonlinear 
electromagnetic theory which does not present scale anomaly.  
If, in addition, the theory is such that $L_F < 0$ 
(that is, if the energy density is positive definite), then 
\begin{equation}
R_{tt} > 0.
\end{equation}
Thus we conclude that Rainich-Wheeler conditions 
for the already unified program characterize 
both linear and nonlinear electromagnetic fields 
as the source of the given gravitational field.

\end{document}